\documentclass[12pt]{article}
\usepackage{a4wide}
\usepackage{graphicx}
\usepackage{axodraw}

\newcommand{\ba}{\begin{eqnarray}}
\newcommand{\ea}{\end{eqnarray}}
\newcommand{\be}{\begin{equation}}
\newcommand{\ee}{\end{equation}}
\newcommand{\tr}{\mbox{tr}}
\newcommand{\re}{\mbox{Re}}
\newcommand{\im}{\mbox{Im}}
\begin{document}
\begin{titlepage}
\begin{flushright}
LU TP 00-17\\
UG-FT/00-114\\
hep-ph/0005189\\
May 2000
\end{flushright}
\begin{center}
{\Large\bf$\mathbf{\varepsilon_K^\prime/\varepsilon_K}$ in the
Chiral Limit\footnote{Partially supported by the European Union TMR Network
EURODAPHNE (Contract No. ERBFMX-CT98-0169).}}\\
\vfill
{\bf Johan Bijnens}\\[0.5cm]
Department of Theoretical Physics 2, Lund University\\
S\"olvegatan 14A, S 22362 Lund, Sweden\\[1cm]
{\bf Joaquim Prades}\\[0.5cm]
Departamento de F\'\i sica Te\'orica y del Cosmos, Universidad de Granada,\\
Campus de Fuente Nueva, E-18002 Granada, Spain
\end{center}
\vfill
\begin{abstract}
The $K\to\pi\pi$ system is analyzed in the chiral limit
within the Standard Model.
We discuss how to connect the short-distance running in the $|\Delta S|=1$
case to the matrix-elements calculated in a low-energy approximation in
a scheme-independent fashion. We calculate this correction
and the resulting Wilson Coefficients.
The matrix elements are calculated to next-to-leading order in the $1/N_c$
expansion and combined with the Wilson coefficients to calculate
the two isospin amplitudes and $\varepsilon_K^\prime$. The $\Delta I=1/2$
rule is reproduced within expected errors and we obtain a substantially
larger value for $\varepsilon_K^\prime/\varepsilon_K$ than most other
analysises. We discuss the reasons for this difference.
We also suggest that the $X$-boson method is an option for
lattice QCD calculations.
\end{abstract}
\vfill
\end{titlepage}

\section{Introduction}

The study of CP-violation took a major step forward in 1999 with
the first direct evidence by two independent experiments for
direct CP-violation\cite{experiment} corroborating the earlier NA31 result.
The average of the two experiments
\be
\left|\frac{\varepsilon_K^\prime}{\varepsilon_K} \right|^{\rm exp}
= (2.13 \pm 0.46) \cdot 10^{-3}\,.
\ee
is higher than the standard calculations obtained \cite{epspeps,RMP} and this
has caused claims for physics beyond the Standard Model,
see e.g. \cite{Burasetc}. The purpose of this paper
is to calculate this quantity within our earlier proposed method
and see whether the above number can be reached or not.
Other opinions critical of this claim are expressed in \cite{RMPtrieste}.

CP-violation was first observed in the sixties and was then shown to proceed
only via $K^0$-$\overline{K^0}$ mixing. This is called indirect CP-violation
and is described by the parameter $\varepsilon_K$
\be
\label{defeps}
\varepsilon_K = \frac{A[K_L\to (\pi\pi)_{I=0}]}{A[K_S\to (\pi\pi)_{I=0}]}\,.
\ee
In addition there is also direct CP-violation in the decay-amplitude
possible in $K\to\pi\pi$ decays. This is parameterized by
\be
\label{defepsp}
\varepsilon'_K = \frac{1}{\sqrt 2}
\left[ \frac{A[K_L\to (\pi\pi)_{I=2}]}{A[K_S\to (\pi\pi)_{I=0}]}
- \frac{A[K_L\to (\pi\pi)_{I=0}]}{A[K_S\to (\pi\pi)_{I=0}]}\, 
\frac{A[K_S\to (\pi\pi)_{I=2}]}{A[K_S\to (\pi\pi)_{I=0}]}
\right] \, .
\ee
Combining both definitions the ratio is directly accessible from
\be
\frac{\varepsilon_K^\prime}{\varepsilon_K} = \frac{1}{\sqrt 2}
\left[ \frac{A[K_L\to (\pi\pi)_{I=2}]}{A[K_L\to (\pi\pi)_{I=0}]}
- \frac{A[K_S\to (\pi\pi)_{I=2}]}{A[K_S\to (\pi\pi)_{I=0}]}
\right]\,.
\ee
This last formula is the basis for the double-ratio method
used in the actual experiments \cite{experiment}.

Theoretical calculations of $\varepsilon_K^\prime/\varepsilon_K$
basically reduce to calculating the $K\to\pi\pi$ amplitudes.
This is in the Standard Model a several step process,
see Section \ref{shortdistance} and e.g. the lectures of \cite{Buras}
for more details. First the heavy particles, top, $Z$ and $W$, are
integrated out, resulting in an effective action in terms
of four-quark operators. Then the renormalization group
evolution equations are used to bring the coefficients in this
effective action down to a hadronic scale thus resumming large logarithms
containing heavy masses. This part has been performed
including gluonic and electroweak Penguin-diagrams
to one-loop in \cite{one-loop} and to two-loops in \cite{two-loop}.

This has then been used with matrix-elements from the lattice
and from the first $1/N_c$-methods \cite{BBG}
to calculate $\varepsilon_K^\prime/\varepsilon_K$
in \cite{epspeps}.
We will not comment further on the lattice calculations. A recent
suggestion about their calculation and more references can be found
in \cite{LL}. We will concentrate on the $1/N_c$ method of calculating
the matrix-elements.

The $1/N_c$-method was first proposed in \cite{BBG} after it was noticed
that using large $N_c$ instead of vacuum-insertion-approximation for
the matrix elements gave a significant improvement for the
$\Delta I=1/2$ rule. They simply identified the cut-off in meson-loops with
the scale in the renormalization group. A better way to provide this
identification was given in \cite{routing,BPBK} by using
colour-singlet bosons to provide the identification of scales.
To one-loop it was proven in \cite{BPBK} that this procedure gives
the correct matching. The work of \cite{Dortmund} is essentially
the continuation of \cite{BBG} using this identification directly
with the output of the renormalization group.

Our work has concentrated on two more improvements. We realized
that our method of using colour-singlet bosons, hereafter referred to
as the $X$-boson method, allowed both a solution to the scale-identification
and to the scheme-dependence that appears at two-loops in the renormalization
group. This was argued in \cite{BPdIhalf} where we used available results
to include part of the needed correction. We proved this explicitly in the
case of $B_K$ in \cite{BPscheme} and below for the $|\Delta S=1|$ case.
The other improvement we have included is the use of the ENJL model for
the couplings of the $X$-bosons to improve on the high-energy behaviour.
For $B_K$ this can be done directly and we presented results
including all quark-mass corrections in \cite{BPBK,BPscheme}.
For $K\to\pi\pi$ decays this produced too unwieldy calculations
and we use instead the method of \cite{kptokpp} to reduce the
calculation to off-shell two-point functions.
Results for the operators $Q_1$ to $Q_6$ in the chiral limit were
presented in \cite{BPdIhalf} and in the same limit for $Q_7$ to $Q_{10}$
below. 

Related work, concentrating on a more QCD inspired improvement of
the $X$-couplings is Ref. \cite{Peris}. We regard this method as very promising
but it has not yet reached the stage where results for the quantities
considered here are possible.

An approach where no attempt at precisely identifying the scales in
the matrix-element and the short-distance running is done is the
chiral quark model approach of \cite{Bertolini}. On the other hand
this series
of papers has attempted to include several effects neglected by
the standard analysises \cite{epspeps} by systematically going to
$p^4$ and thus provided an indication that the errors
in \cite{epspeps} were severely underestimated.

Another approach to determining the non-leptonic matrix-elements
is to use QCD sum rules to calculate more inclusive quantities
and then relate those to the matrix-elements. Examples of this approach
are \cite{JP}. 

As discussed in \cite{BPBK} for the case of the $B_K$
parameter, a general problem with this method is the existence
of operators in the CHPT description that contribute to the inclusive
quantities but not to the decays $K\to\pi\pi$. The reason is that
the so-called spurion-fields have now non-vanishing derivatives
\footnote{In the notation of Section \ref{CHPT} an example of an extra
octet term is 
$\tr(\Delta_{32}u_\mu)\tr(\nabla^\mu(\Delta_{11}+\Delta_{22}+\Delta_{33}))$.}.

We would like to stress that our method basically reproduces the
$\Delta I=1/2$ rule, has no free input parameters as several other
models have and includes a correct scheme and 
scale identification at all stages.
As will become obvious, we also reproduce the value of 
$\varepsilon_K^\prime/\varepsilon_K$ within the uncertainty of our
approach.

Our $X$-boson method can be used in general and is not restricted to
$1/N_c$. E.g., by adapting the method of \cite{Eichten} to
use $X$-bosons instead of photons it might provide an alternative
lattice QCD method to calculate $K\to\pi\pi$ on the lattice.

In the next section we introduce the basic notation used
to describe $K\to\pi\pi$ decays to lowest order in Chiral
Perturbation Theory (CHPT). Section \ref{shortdistance} describes the
$X$-boson method and calculates the corrections needed to
identify the scales in QCD and the low-energy model used.
The main result is presented in the appendix.
Section \ref{longdistance} presents the calculation in $1/N_c$
using CHPT and the ENJL model for the $X$-boson couplings for the
electroweak Penguin operators. The numerical input used is
summarized in Section \ref{numinput}.
The results from the previous sections and the matrix-elements
of \cite{BPdIhalf} are then combined to calculate the parameters
in the CHPT $\Delta S=1$ Lagrangian in Section \ref{resultsGi} and
presented using the more standard bag parameters, $B_i$ in
Section \ref{Bi}. The next section contains the calculation
of $\varepsilon_K^\prime$ and briefly of $\varepsilon_K$ as well.
The last section discusses the result for $\varepsilon_K^\prime/\varepsilon_K$
and adds the two corrections which are expected to dominate.

\section{Chiral Perturbation Theory Lagrangian}
\label{CHPT}

We use the standard Chiral Perturbation Theory (CHPT)
counting with $p$ a momentum, energy or
meson-mass and $e$ the electromagnetic coupling constant $e=\sqrt{4\pi\alpha}$.
The lowest order strong and electromagnetic Lagrangian is given by
\be
\label{lagstrong}
{\cal L}^{(2)} =
\frac{F_0^2}{4}\tr\left(u_\mu u^\mu+\chi_+\right)
+ e^2 \tilde C_2 \tr\left(Q U Q U^\dagger\right)
\ee
with $U = u^2 = \exp(\lambda^a\pi^a/F_0)$ and $u_\mu = i u^\dagger D_\mu U
u^\dagger$. $\lambda^a$ are the Gell-Mann matrices and the $\pi^a$
are the pseudoscalar-mesons $\pi$, $K$, and $\eta$.
$Q=\mbox{diag}(2/3,-1/3,-1/3)$ is the light-quark-charge matrix and
$\chi_+ = u^\dagger\chi u^\dagger + u \chi^\dagger u$.
$\chi = 2 B_0 \, \mbox{diag}(m_u,m_d,m_s)$ collects the light-quark 
masses. To this order $F_0=F_\pi$ is the pion decay coupling constant. 
An introduction to CHPT can be found in the lectures \cite{CHPTlectures}.

To order $e^0 p^2$ and $e^2 p^0$, the chiral Lagrangian 
describing $|\Delta S|=1$ transitions is given by
\ba
\label{lagdS1}
{\cal L}^{(2)}_{|\Delta S|=1}&=&
C F_0^6 e^2 G_E \,  \tr\left(\Delta_{32}\tilde{Q}\right)
+ C F_0^4 \Bigg[G_8\tr\left(\Delta_{32}u_\mu u^\mu\right)
+G_8^\prime\tr\left(\Delta_{32}\chi_+\right) 
\nonumber \\
&+&G_{27}t^{ij,kl}\tr\left(\Delta_{ij}U_\mu\right)
  \tr\left(\Delta_{kl}u^\mu\right)\Bigg] +\mbox{h.c.}
\ea
with $\Delta_{ij} = u\lambda_{ij}u^\dagger$,
$(\lambda_{ij})_{ab} = \delta_{ia}\delta_{jb}$,
$\tilde{Q}=u^\dagger Q u$; 
\ba
C&=& -\frac{3}{5} \frac{G_F}{\sqrt 2} \, V_{ud} V_{us}^* \approx
-1.06 \cdot 10^{-6} \, {\rm GeV}^{-2} \, . 
\ea
The SU(3) $\times$ SU(3) tensor
$t^{ij,kl}$ can be found in \cite{kptokpp}.

The $K\to \pi \pi$ invariant amplitudes can be decomposed into definite
isospin quantum numbers amplitudes as $[A\equiv-i T]$
\ba
A[K_S\to \pi^0\pi^0] &\equiv& \sqrt{\frac{2}{3}} \, A_0
-\frac{2}{\sqrt 3} \, A_2 \, , \nonumber \\
A[K_S\to \pi^+\pi^-] &\equiv& \sqrt{\frac{2}{3}} \, A_0
+\frac{1}{\sqrt 3} \, A_2 \, , \nonumber \\
A[K^+\to\pi^+\pi^0] &\equiv& \frac{\sqrt 3}{2} \, A_2 \, . 
\ea
Where $K_S\simeq K_1^0 + 
\varepsilon_K K_2^0$; 
 $K_{1(2)}^0 \equiv (K^0-(+){\overline K^0})/\sqrt 2$,   and
CP$(K^0_{1(2)})=+(-)K^0_{1(2)}$.  The final state 
interaction phases $\delta_0$ and $\delta_2$ are included
into the amplitudes $A_0$ and $A_2$  as follows.
\ba
A_0&\equiv& -i \, a_0 \, e^{i\delta_0}
\quad\quad\mbox{for the isospin $1/2$ amplitude;}
\nonumber\\
A_2&\equiv& -i \, a_2 \,
e^{i\delta_2}\quad\quad\mbox{for the isospin $3/2$ amplitude} \, .
\ea
Using the Lagrangian of Eq. (\ref{lagdS1}) we obtain
\ba
a_0 & = &\frac{\sqrt 6}{9} \, C F_0\left[\left(9G_8+G_{27}\right) 
(m_K^2-m_\pi^2) - 6 e^2 G_E F_0^2 \right]
\nonumber\\
a_2 & = & \frac{\sqrt 3}{9} \, C F_0\left[10 G_{27} \, 
(m_K^2-m_\pi^2) - 6 e^2 G_E F_0^2 \right]\,.
\ea
In the presence of CP-violation the couplings $G_8$, $G_{27}$, and
$G_E$ get an imaginary part. In the Standard Model,
$\im \, G_{27}$ vanishes and
$\im \, G_8$ and $\im \, G_E$ are proportional
to $\im \, \tau$ with
 $\tau \equiv -\lambda_t /\lambda_u$ and $\lambda_i\equiv
V_{id} V_{is}^*$.

\section{Short-Distance}
\label{shortdistance}

In Section \ref{longdistance} we will calculate the long-distance contributions
in the $X$-boson scheme. This scheme requires only the matching of currents
between the low-energy model/theory and the short-distance QCD calculations.
This way the scheme- and scale-dependence inherent in most of the
other calculations is consistently avoided. This scheme has been shown
to one-loop to reproduce the correct matching \cite{BPBK}
and was argued to work to all orders in \cite{BPdIhalf}. The precise
proof to two-loops was done in \cite{BPscheme} for the case of the
$B_K$-parameter. Exactly the same calculation can be performed for
the $|\Delta S=1|$ case. There is no new problem but in
practice the calculation is much longer because there are now ten operators
instead of one. Penguin diagrams need to be taken into account and
photonic loops as well as gluonic ones have to be considered.
We will therefore present fewer explicit expressions than in \cite{BPscheme}.

We start from the effective action that is the output of the
renormalization group running at two-loops which is given by
\be
\label{effactionOPE}
\Gamma_{\mbox{\small eff}} =
-\frac{G_F}{\sqrt{2}}V_{ud} V_{us}^*\int d^4x \sum_{i=1,10} C_i Q_i(x)\,,
\ee
with $C_i= z_i+y_i\tau$ and
\ba
\label{defQi}
Q_1 &=& (\bar s_\alpha\gamma_\mu u_\beta)_L (\bar u_\beta\gamma^\mu d_\alpha)_L
\nonumber\\
Q_2 &=& (\bar s_\alpha\gamma_\mu u_\alpha)_L (\bar u_\beta\gamma^\mu d_\beta)_L
\nonumber\\
Q_3 &=& (\bar s_\alpha\gamma_\mu d_\alpha)_L
\sum_{q=u,d,s} (\bar q_\beta\gamma^\mu q_\beta)_L
\nonumber\\
Q_4 &=& (\bar s_\alpha\gamma_\mu d_\beta)_L
\sum_{q=u,d,s} (\bar q_\beta\gamma^\mu q_\alpha)_L
\nonumber\\
Q_5 &=& (\bar s_\alpha\gamma_\mu d_\alpha)_L
\sum_{q=u,d,s} (\bar q_\beta\gamma^\mu q_\beta)_R
\nonumber\\
Q_6 &=& (\bar s_\alpha\gamma_\mu d_\beta)_L
\sum_{q=u,d,s} (\bar q_\beta\gamma^\mu q_\alpha)_R
\nonumber\\
Q_7 &=& (\bar s_\alpha\gamma_\mu d_\alpha)_L
\sum_{q=u,d,s} \frac{3}{2}e_q(\bar q_\beta\gamma^\mu q_\beta)_R
\nonumber\\
Q_8 &=& (\bar s_\alpha\gamma_\mu d_\beta)_L
\sum_{q=u,d,s} \frac{3}{2}e_q(\bar q_\beta\gamma^\mu q_\alpha)_R
\nonumber\\
Q_9 &=& (\bar s_\alpha\gamma_\mu d_\alpha)_L
\sum_{q=u,d,s} \frac{3}{2}e_q(\bar q_\beta\gamma^\mu q_\beta)_L
\nonumber\\
Q_{10} &=& (\bar s_\alpha\gamma_\mu d_\beta)_L
\sum_{q=u,d,s} \frac{3}{2}e_q(\bar q_\beta\gamma^\mu q_\alpha)_L
\ea
with $(\bar q\gamma_\mu q^\prime)_{(L,R)} 
= \bar q \gamma_\mu(1\mp\gamma_5)q^\prime$;
$\alpha$ and $\beta$ are colour indices.

The Wilson coefficients 
$C_i(\mu_R)$ are dependent on the scheme chosen for $\gamma_5$,
the choice of evanescent operators, the scale $\mu_R$ in the
renormalization group as well as other scheme choices like 
the choice of infrared regulators.
All of them have to be consistently treated when calculating matrix
elements in order to reach a physical result.
 Reviews of this can be found in \cite{RMP} and \cite{Buras}.
The main original references are \cite{two-loop}.
We have chosen two slightly different implementations of the running, both
in the NDR scheme. One where we linearize fully the NLO part, this way
all divergences appearing in the expressions can be analytically
resolved\footnote{Note that in none of the papers of \cite{two-loop}
all necessary formulae were given, we have worked them out ourselves.}
and the scheme dependence is canceled exactly except for effects
of order $\alpha_S(m_c)^2$. The other implementation is where we
exactly solve the two-loop evolution equations for the $C_i(\mu_R)$.
 In this
case the scheme-dependence is only canceled to order $\alpha_S(\mu_R)^2$.
We regard the difference between these two as a rough estimate of the
uncertainty due to higher order corrections in $\alpha_S(\mu_R)$.

The next step is now to replace the effective action (\ref{effactionOPE})
by another equivalent effective action in $D=4$
where the quarks only appear  in vector, axial-vector currents, and  
scalar or pseudoscalar densities.  These we know how to hadronize.
Specifically we replace Eq. (\ref{effactionOPE}) by
\ba
\label{effactionX}
\Gamma_{\mbox{\small X}}
&=&g_1 X_1^\mu \left((\bar s\gamma_\mu d)_L + (\bar u\gamma_\mu u)_L\right)
+ g_2 X_2^\mu  \left((\bar s\gamma_\mu u)_L + (\bar u\gamma_\mu d)_L\right)
\nonumber\\&&
+g_3 X_3^\mu \left((\bar s\gamma_\mu d)_L + 
\sum_{q=u,d,s}(\bar q\gamma_\mu q)_L\right)
+g_4 \sum_{q=u,d,s} X_{q,4}^\mu 
    \left((\bar s\gamma_\mu q)_L+(\bar q\gamma_\mu q)_L\right)
\nonumber\\&&
+g_5 X_5^\mu \left((\bar s\gamma_\mu d)_L + 
\sum_{q=u,d,s}(\bar q\gamma_\mu q)_R\right)
+g_6 \sum_{q=u,d,s} X_{q,6} \left((\bar s  q)_L + 
(-2)(\bar q d)_R\right)
\nonumber\\&&
+g_7 X_7^\mu \left((\bar s\gamma_\mu d)_L + 
\sum_{q=u,d,s}\frac{3}{2}e_q(\bar q\gamma_\mu q)_R\right)
+g_8 \sum_{q=u,d,s} X_{q,8} \left((\bar s  q)_L + 
(-2)\frac{3}{2}e_q(\bar q d)_R\right)
\nonumber\\&&
+g_9 X_9^\mu \left((\bar s\gamma_\mu d)_L + 
\sum_{q=u,d,s}\frac{3}{2}e_q(\bar q\gamma_\mu q)_L\right)
+g_{10} \sum_{q=u,d,s} X_{q,10}^\mu 
    \left((\bar s\gamma_\mu q)_L+
\frac{3}{2}e_q(\bar q\gamma_\mu q)_L\right).
\nonumber\\
\ea
Here all colour sums are performed implicitly inside the brackets
and $(\bar q q^\prime)_{(L,R)} 
= \bar q (1\mp\gamma_5)q^\prime$.

For simplicity we choose all $X$-bosons to have the same mass.
We now determine the couplings $g_i$ as a function of the $C_i$
by taking matrix elements of both sides between quark and
and gluon external states, labelled by $\psi$ and $\psi^\prime$
and require
\be
\label{match}
\langle\psi^\prime|e^{i\Gamma_{\mbox{\tiny eff}}}|\psi\rangle =
\langle\psi^\prime|e^{i\Gamma_{\mbox{\tiny X}}}|\psi\rangle +
{\cal O}(1/M_X^4)\,.
\ee
Both the left- and right-hand-side can be written in terms of the
tree level matrix-elements of the $Q_i$ between the states $\psi,\psi^\prime$.
Leading to
\ba
\label{match2}
\hskip-1cm
&&\sum_{j} \left[\delta_{ij}+\frac{\alpha_s(\mu_R)}{\pi}
 \left( \gamma_{ij}  \log\frac{m}{\mu_R}
+r_{ij}+F_{ij}(m^2,\psi,\psi^\prime)\right) \right. 
\nonumber \\ 
&+&\frac{\alpha}{\pi}
\left. 
\left(\gamma^e_{ij}  \log\frac{m}{\mu_R}
+r^e_{ij} +F^e_{ij}(m^2,\psi,\psi^\prime)
\right)
\right] C_j(\mu_R)
\nonumber\\&=&
\sum_{j} \left[\delta_{ij}+ \frac{\alpha_s(\mu_C)}{\pi} \left(
\left(\gamma_{ij}-\tilde \gamma_{ij}\right)
\log\frac{m}{M_X} +\tilde \gamma_{ij} \log \frac{m}{\mu_C}
+\tilde r_{ij}+F_{ij}(m^2,\psi,\psi^\prime)\right) \right. \nonumber \\
&+&\frac{\alpha}{\pi} \left. 
\left(\left(\gamma^e_{ij}-\tilde \gamma^e_{ij}\right) \log{\frac{m}{M_X}} 
+ \tilde \gamma^e_{ij}  \log \frac{m}{\mu_C}
+\tilde r^e_{ij} +F^e_{ij}(m^2,\psi,\psi^\prime)
\right)
\right] \frac{g_j(\mu_C)^2}{M_X^2}
\ea
The parameter $m$ is some IR regulator.
For conserved currents $\tilde \gamma_{ij} =0$ and there are no
 logarithms of the scale $\mu_C$.  For non-conserved currents or densities
$\tilde \gamma_{ij}$ can be related directly to their anomalous dimensions.

The infrared sensitive parts are in the functions
$F_{ij}$ and in $F_{ij}^e$ and
are identical on both sides. We have checked this explicitly for
a set of states similar to the one chosen in \cite{BPscheme}.
The resulting formulas are very long and only reinforce the argument for
complete infrared ambiguity cancellation.
Notice that the matrices $\gamma_{ij}$ and
$\gamma_{ij}^e$
are one-loop anomalous dimension matrices
of the set of four-quark operators (\ref{defQi}) \cite{one-loop}.
The left-hand-side is thus completely scheme-independent
to order $\alpha_S^2$. The scheme-dependence of the $C_j(\mu_R)$
is canceled by the scheme-dependence of the constants
$r_{ij}$ and $r_{ij}^e$.
The right-hand-side is similarly scale-independent

{}From Eq. (\ref{match2}) we can now obtain
\ba
\label{match3}
\hskip-1cm
 \frac{g_i(\mu_C)^2}{M_X^2}&=&
\sum_j \left[\delta_{ij}+
\frac{\alpha_s(\mu_C)}{\pi }\left(\gamma_{ij}\log\frac{M_X}{\mu_R} 
+ \tilde \gamma_{ij} \log\frac{\mu_C}{M_X}
+r_{ij}-\tilde r_{ij}\right) \right. \nonumber \\
& &+ \frac{\alpha}{\pi}\left(\gamma^e_{ij}\log\frac{M_X}{\mu_R}
+ \tilde \gamma^e _{ij} \log\frac{\mu_C}{M_X}
+ r^e_{ij}-\tilde r^e_{ij}\right) 
\nonumber \\
&&\left. +  O\left(\alpha_S(\mu_R)-\alpha_S(\mu_C)\right)
\right]C_j(\mu_R)\,.
\ea
The $X$-boson couplings are now completely scheme- and scale-independent
and the large logarithms of $M_W$ are resummed and included in the values
of the couplings $g_i$.
The explicit expressions for the differences $r_{ij}-\tilde r_{ij}$
and $r^e_{ij}-\tilde r^e_{ij}$ are given in the appendix.

We have not used any large $N_c$ arguments up to this point. Now we need
to calculate the matrix-elements of $X$-boson exchange between hadronic
external states. As described in more detail in \cite{BPscheme} and references
therein, we do this by rotating the integral over the $X$-boson momenta
into Euclidean space and splitting it into two-parts separated by an
Euclidean cut-off $\mu$.

The large momentum part of this integral [between $\mu$ and $\infty$]
can be simply calculated. The large
$X$-boson momentum must flow back via quark- or gluon-lines and can be
calculated perturbatively. The part where the momentum flow-back is through
the hadronic wave-functions is suppressed by $1/\mu^2$ and
can be neglected. Precisely as described in \cite{BPscheme}
the result is proportional to $\alpha_S(\mu)$ and is thus suppressed
by $1/N_c$ compared to the tree-level matrix elements
or it is down by an extra factor of $\alpha$ and reads
\ba
\langle out | X_j\mbox{-exchange} | in\rangle^{\mbox{\tiny Short-Distance}}&=&
\sum_i\left[
\frac{\alpha_S(\mu)}{\pi}\left(\gamma_{ij}-\tilde \gamma_{ij}\right)
+\frac{\alpha}{\pi}\left(\gamma^e_{ij}-\tilde \gamma^e_{ij}\right)
\right]
\nonumber\\&\times&
\log{\frac{M_X^2}{\mu^2}} \, \, 
\langle out | X_i\mbox{-exchange}
 | in\rangle_{\mbox{\tiny Leading $1/N_c$}}\,.
\ea
This part of the integration can thus be included by replacing
in Eq. (\ref{match3}) everywhere $\log M_X$ by $\log \mu$ leading to
\ba
\label{match4}
\hskip-1cm
\left( \frac{g_i(\mu_C,\mu)^2}{M_X^2}\right)_{\mbox{eff}}&=&
\sum_j \left[\delta_{ij}+
\frac{\alpha_s(\mu_C)}{\pi }\left(\gamma_{ij}\log\frac{\mu}{\mu_R} 
+ \tilde \gamma_{ij} \log\frac{\mu_C}{\mu}
+r_{ij}-\tilde r_{ij}\right) \right. \nonumber \\
& &+ \frac{\alpha}{\pi}\left(\gamma^e_{ij}\log\frac{\mu}{\mu_R}
+ \tilde \gamma^e _{ij} \log\frac{\mu_C}{\mu}
+ r^e_{ij}-\tilde r^e_{ij}\right) \nonumber \\
&&\left. +  O\left(\alpha_S(\mu_R)-\alpha_S(\mu_C)\right)
\right]C_j(\mu_R)\,.
\ea
It is obvious from this that the final answer for the decays will not
depend on the artificial choice of the mass $M_X$.

The low energy part of the integral [between 0 and $\mu$]
 over the $X$-boson momenta must now
be performed in a particular low-energy approximation.
We must therefore identify correctly the vector, axial-vector currents
and scalar and pseudoscalar densities appearing in Eq. (\ref{effactionX})
in the low-energy model. This is no problem for the vector and axial-vector
currents since they are currents from a conserved symmetry.
For the scalar and pseudoscalar densities the situation
is somewhat more uncertain.
There is  remaining dependence on $\log\mu_C$
from the terms with $\tilde\gamma_{ij}$ and $\tilde\gamma^e_{ij}$.
The scale $\mu_C$ is the one at which currents and densities 
in QCD and in the low-energy model (CHPT and/or ENJL) are matched.
We stress again that this problem of matching of the (pseudo)scalar densities
to the low-energy model, though uncertain, is a much more tractable problem
than the original matching of the four-quark operators
of Eq. (\ref{effactionOPE}) directly. Practically, 
we do this matching when
substituting the CHPT or ENJL quark-condensate by the QCD
value at $\mu_C \simeq 1$~GeV.
For other scales  we run the quark density with its QCD anomalous dimension. 

\section{Long-Distance}
\label{longdistance}

As described before in \cite{BPBK,BPdIhalf} we can calculate the
very low-energy part using CHPT to lowest order for the $X$-boson
couplings.\footnote{Note that we use lowest order CHPT in two different
meanings. Lowest order in CHPT in the couplings in the Lagrangian
of Eq. (\ref{lagdS1}) and lowest order in CHPT in the $X$-boson couplings
which are used to calculate the coefficients in Eq. (\ref{lagdS1}).}
For the intermediate energy region 
one can use hadronic models for the $X$-boson couplings.
They must have the correct chiral
symmetry and reproduce  at very low energy the model-independent CHPT results.
Here we will use the ENJL model. As we noted earlier in
\cite{BPdIhalf,BPscheme} this provides a substantial improvement over
using lowest-order CHPT $X$-boson couplings.
It brings in {\em no new}
undetermined parameters, describes a fairly wide range of
low-energy hadronic phenomena \cite{ENJL} and satisfies some
short-distance constraints from QCD as well \cite{ENJL2point}.

The contributions to order $e^0 p^2$ were already
worked out in \cite{BPdIhalf}. We now add the results relevant
to order $e^2 p^0$.
They can be calculated using our earlier methods, a well-defined
off-shell $K$-$\pi$ Green function \cite{kptokpp}, or with a reduction method
as was used in \cite{Peris}.
The comparison between both and a more extensive
discussion of the results in this section will be
presented elsewhere \cite{BGP}.

The coupling $G_E F_0^6$ has only contributions
from $Q_7$ and $Q_8$. The Wilson coefficients $C_7$ is order 1
in $1/N_c$ and $C_8$ is order $1/N_c$.  

\begin{figure}
\begin{center}
\setlength{\unitlength}{1.5pt}
\begin{picture}(210,40)(-10,-10)
\SetScale{1.5}
\Line(0,0)(30,0)
\DashLine(30,0)(60,0){5}
\Line(60,0)(90,0)
\Text(0,0)[]{$\bigotimes$}
\Vertex(30,0){3}
\Vertex(60,0){3}
\Text(90,0)[]{$\bigotimes$}

\Line(110,0)(190,0)
\DashLine(150,0)(150,30){5}
\Text(110,0)[]{$\bigotimes$}
\Text(190,0)[]{$\bigotimes$}
\Vertex(150,0){3}
\Vertex(150,30){3}
\end{picture}
\end{center}
\caption{\label{figdiagLO} The lowest order, $e^2p^0$, diagrams for the
contributions from $Q_8$. $\bigotimes$ is an insertion of the
pseudo-scalar currents. The dashed line is an $X$-boson and the dots
are the $X$-boson-pseudoscalar couplings.
Full lines are pseudo-scalar propagators.}
\end{figure}
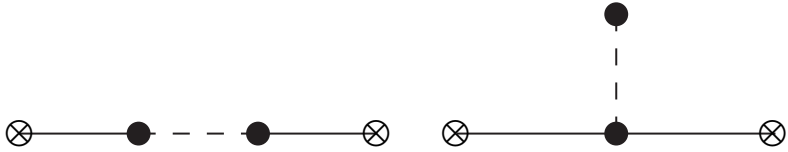

The leading in $1/N_c$ contribution
to the matrix element of $Q_8$
comes from the two diagrams in Figure \ref{figdiagLO}
and is of order $N_c^2$.
Using quark-loop diagrams one can prove model-independently \cite{BGP}
that in the chiral limit the next-to-leading order in $1/N_c$ 
vanishes to all orders in $X$-boson couplings.
The non-factorizable contributions from $Q_8$
thus only start at order $e^2 p^2$ and do not contribute to $G_E$. 
This contribution leads to $G_E F_0^6$ of order $N_c$.

The leading contribution in $1/N_c$ from $Q_7$
is factorizable but is only of order $e^2 p^2$.
The next-to-leading contribution in $1/N_c$ contributes
to order $e^2p^0$ at order $N_c$ to the matrix-element and
leads again to $G_E F_0^6$ of order $N_c$. The $N_c$ counting presented
here is scheme-independent.

We can now calculate the leading contributions using lowest-order CHPT for the
$X$-boson couplings and obtain
\be
\label{geq8}
e^2 G_E[Q_8] = -5\frac{B_0^2(\mu)}{F_0^2}C_8(\mu) =
-5 \frac{\langle 0 | \overline q q |0\rangle^2(\mu)}{F_0^6} \, C_8(\mu)
\ee
where the Wilson coefficient is the one appropriate for our $X$-boson
scheme as discussed in Section \ref{shortdistance}
and $F_0^2 \, B_0(\mu)\equiv - \langle 0 | \overline q q | 0\rangle (\mu)$
the quark condensate in the chiral limit. There are also $O(e^2)$
contributions from $Q_8$ to $G_8$ and $G_{27}$.

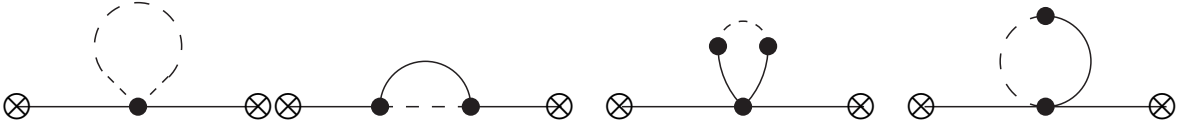
\begin{figure}
\begin{center}
\setlength{\unitlength}{1.14pt}
\begin{picture}(385,100)(10,-10)
\SetScale{1.14}
\Line(10,0)(90,0)
\DashCArc(50,20)(14.414,315,215){5}
\DashLine(50,0)(40,10){3}
\DashLine(50,0)(60,10){3}
\Vertex(50,0){3}
\Text(10,0)[]{$\bigotimes$}
\Text(90,0)[]{$\bigotimes$}

\Line(100,0)(130,0)
\DashLine(130,0)(160,0){5}
\CArc(145,0)(15,0,180)
\Line(160,0)(190,0)
\Text(100,0)[]{$\bigotimes$}
\Vertex(130,0){3}
\Vertex(160,0){3}
\Text(190,0)[]{$\bigotimes$}

\Line(210,0)(290,0)
\CArc(270,20)(28.284,180,225)
\CArc(230,20)(28.284,315,360)
\DashCArc(250,20)(8.284,0,180){3}
\Vertex(258.284,20){3}
\Vertex(241.716,20){3}
\Vertex(250,0){3}
\Text(210,0)[]{$\bigotimes$}
\Text(290,0)[]{$\bigotimes$}

\Line(310,0)(390,0)
\CArc(350,15)(15,270,90)
\DashCArc(350,15)(15,90,270){5}
\Vertex(350,0){3}
\Vertex(350,30){3}
\Text(310,0)[]{$\bigotimes$}
\Text(390,0)[]{$\bigotimes$}
\end{picture}
\end{center}
\caption{\label{figdiagNLO} The lowest order, $e^2p^0$, diagrams for the
contributions from $Q_7$ but next-to-leading in $1/N_c$.
Notation is as in Fig. \ref{figdiagLO}.}
\end{figure}

The calculations from the diagrams in Fig. \ref{figdiagNLO} lead to
\be
\label{geq7}
e^2 G_E[Q_7] = -\frac{15}{2} \, 
\frac{\mu^4}{16\pi^2 F_0^4}  C_7(\mu)\,.
\ee
Both  results in Eq. (\ref{geq8}) and (\ref{geq7}) agree with
those of \cite{Peris} when the results for $e^2 G_E[Q_7]$ listed there
are restricted to lowest order CHPT only. Notice that the formula (20)
in that reference is only valid for $\Lambda^2 >> M_A^2, M_V^2$
and can only be applied at much higher scales. 
That approximation partially explains the
large scale dependence found below  1.3 GeV in \cite{Peris}.
$\Lambda$ there corresponds to $\mu$ here.

We stress again that
the Wilson coefficients in (\ref{geq8}) and (\ref{geq7}) are
in the $X$-boson scheme where the matrix-elements can be identified
unambiguously.

We can now turn to the ENJL model for the $X$-boson couplings.
These results have a better high energy
behaviour. As mentioned already above, for $Q_8$ the next-to-leading order
in $1/N_c$ vanishes
and the ENJL result thus coincides
with Eq. (\ref{geq8}). The result for $e^2 G_E[Q_7](\mu)/C_7(\mu)$ is given in
Table \ref{tabGEQ7}.
The numbers should be compared with the lowest order result
for $e^2 G_E[Q_8]^{\mbox{\small ENJL}}=-5040 \, C_8(\mu)$ using 
$B_0^{\mbox{\small ENJL}}=2.8$~GeV.
The latter needs to be corrected for the QCD value of 
$B_0(1~{\rm GeV})=1.75$~GeV \cite{BPR} and QCD running.
This correction is the one discussed in Section \ref{shortdistance}
in order to identify currents and densities
in QCD and in the low-energy model and is unambiguous.
The usual assumption on the value of $m_s$ needed shows up in
our way in the value of $B_0$, i.e. the quark condensate in the chiral limit.
The relevant quantity is really the value of the quark condensate\footnote{
In fact, the next-to-leading order CHPT
corrections proportional to quark masses are not the ones
necessary to change
$\langle 0| \overline q q | 0 \rangle$ into
$\langle 0| \overline s s | 0 \rangle$ \cite{BPdIhalf,PR99}.
So the matrix element of $Q_6$
and $Q_8$ at higher orders
are not proportional to the strange quark condensate or
the strange quark mass via PCAC. 
Interpreting the value of $B_0$ in terms of $m_s$ via lowest order 
PCAC as done in the usual $\varepsilon^\prime_K/\varepsilon_K$ analyses it
corresponds to $m_s(1~\mbox{GeV}) \approx 140$~MeV.}
rather than the value of $m_s$.
\begin{table}
\begin{small}
\begin{tabular}{c|*{11}{c}}
$\mu$  &
 0.10  &
 0.20  &
 0.30  &
 0.40  &
 0.50  &
 0.60  &
 0.70  &
 0.80  &
 0.90  &
 1.00 \\
\hline
  $\frac{e^2 G_E[Q_7](\mu)}{C_7(\mu)}$ &
$   -0.1$& 
$   -1.5$& 
$   -6.9$& 
$  -19.6$& 
$  -42.5$& 
$  -77.5$& 
$ -126$  &
$ -189$  &
$ -266$  &
$ -358$  
\end{tabular}
\caption{\label{tabGEQ7} The ENJL model results for the contribution
from $Q_7$ to $e^2 G_E(\mu)$ in the $X$-boson scheme.}
\end{small}

\end{table}

\section{Numerical Input}
\label{numinput}

The input values we use are
\ba
\overline m_t(m_t)= (165 \pm 5) ~\mbox{GeV} &\displaystyle M_W = 80.3945 
~\mbox{GeV}& M_Z=91.1872 ~\mbox{GeV}\nonumber\\
sin^2(\theta_W) = 0.2315 &\displaystyle \overline m_b(m_b) = 4.4~\mbox{GeV} &
\overline m_c(m_c) = 1.23~\mbox{GeV} \nonumber\\
|V_{us}|=0.2196 &\displaystyle |V_{cb}|=0.040\pm0.002 &
|V_{ub}/V_{cb}| = 0.090\pm0.025
\ea
Quark masses are in the $\overline {\rm MS}$ scheme, $M_Z$ and $M_W$
are pole masses. 
The electroweak input values have been taken from the review
by Sirlin \cite{Sirlin}, the CKM matrix values from the review by
Falk \cite{Falk}.

We will simply take $\sin (\delta)=1$
so that
\be
\im \, \tau = -6.72 \cdot 10^{-4}\,.
\ee

In addition we use two sets of values for the strong coupling constant.
Set I we choose with the central value of the measurement
at the $\tau$-mass, $\alpha_S(m_\tau)=0.345$ \cite{Pich}
and set II with the world average value at $M_Z$, $\alpha_S(M_Z)=0.1186$
\cite{Sirlin}. We use exact running with the two-loop beta function,
this is what is most closely needed for the running of the Wilson
coefficients.

The scale-independent
coefficients $\hat \eta_{i=1,3}$ needed for the calculation of $\varepsilon_K$
we take from \cite{HerrlichNierste}, 
for $\hat \eta_1$ we interpolate the values
given in Table 5 of that reference and obtain
$\hat \eta_1=1.93$ (Set I) and $\hat \eta_1=1.53$ (set II). The others are
$\hat \eta_2=0.57$ and $\hat \eta_3=0.47$

In addition we use $F_K=112.7$ MeV, $\hat B_K = 0.77\pm0.10$ \cite{BPscheme},
$F_0 = 87$~MeV, $B_0^{\mbox{\tiny QCD}}(1{\rm~GeV})$ $= 1.75$~GeV and the
parameters of the ENJL model as determined by the fit in \cite{ENJL}.

\section{Results for $G_8$, $G_{27}$ and $G_E$}
\label{resultsGi}

We now combine the matrix-elements and the Wilson coefficients in the
$X$-boson scheme to obtain the quantities needed for the
$\Delta I=1/2$ rule and $\varepsilon_K'/\varepsilon_K$.

We use here the input values given above and the two sets of values
for $\alpha_S$, labelled I and II. In order to have some estimate
of higher order corrections in $\alpha_S$ we have treated the NLO running
of the Wilson coefficients and the transition to the $X$-boson scheme
in two ways which only differ in higher orders in $\alpha_S$. One
where we use the exact solution of the two-loop evolution equations
and multiply with the corrections to go from the NDR to the $X$-boson-scheme.
This is labelled with {\em mul} for multiplicative in the figures.
The other option is to only approximately solve the NLO running as a series
in $\alpha_S$. We also include the scheme-correction now additively
with the running. This is labelled {\em add} in the figures.
Other estimates give a similar variation.

For definiteness we also set all the three scales $\mu_R$, $\mu_C$
and $\mu$ equal to each other. The variation with this assumption for $B_K$
was shown in \cite{BPscheme} and is smaller than the difference between the
two values of $\alpha_S$ used.

\begin{figure}
\includegraphics[height=0.49\textwidth,angle=270]{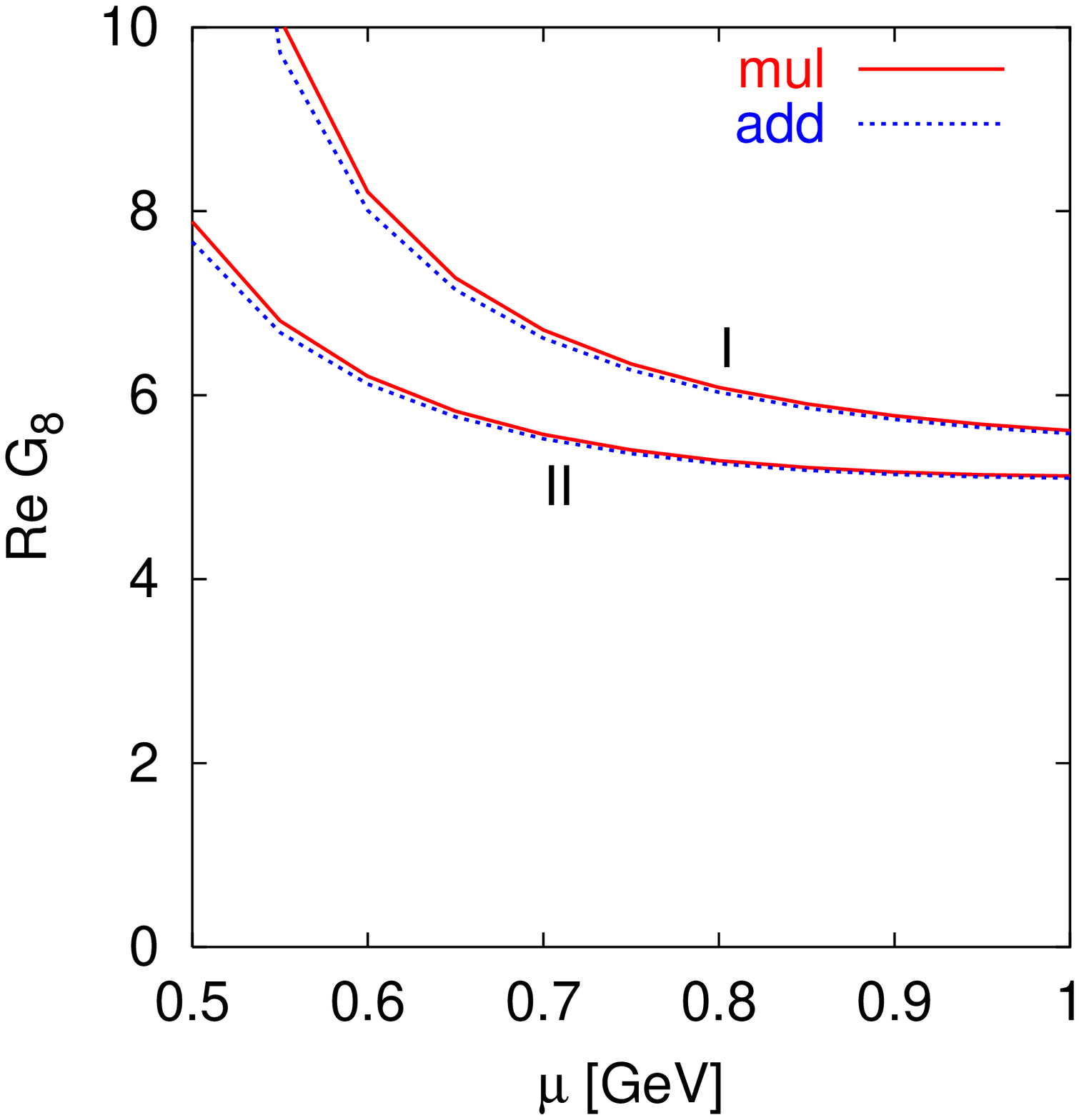}
\includegraphics[height=0.49\textwidth,angle=270]{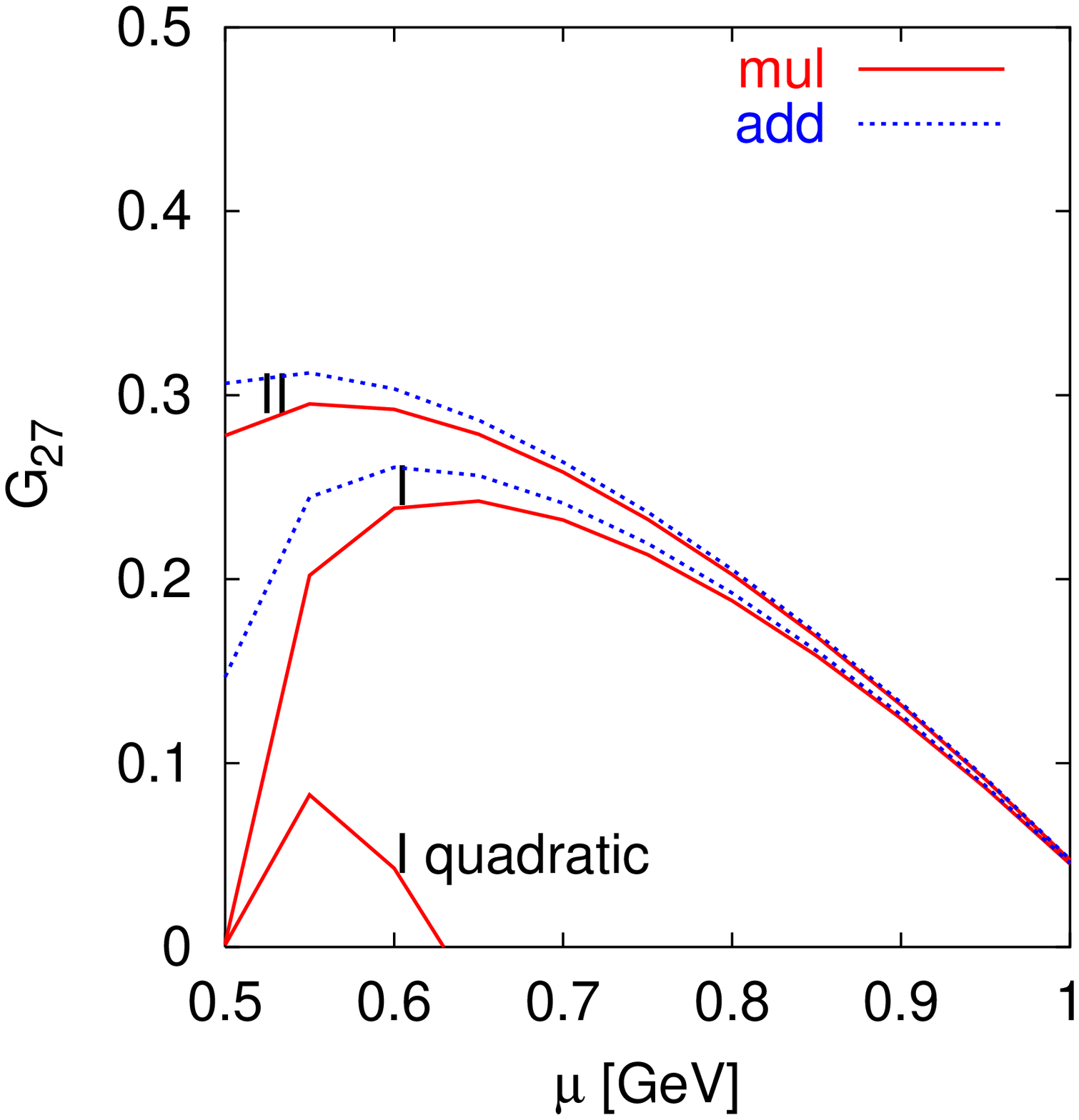}
\caption{\label{figdIhalf} The real part of $G_8$
and $G_{27}$ as a function of $\mu$. The experimental value
of $\re \, G_8$ is about 6.2 and for $G_{27}$ about 0.48.
I and II refer to the two values of $\alpha_S$ used and the two curves
to two ways of solving the evolution equations as described in the text.
In the large $N_c$ limit both $G_{27}$ and $\re G_8$ are equal to 1.}
\end{figure}

The results relevant for the $\Delta I=1/2$ rule are shown
in Fig. \ref{figdIhalf}.
For $ \re \, G_8$ we obtain a reasonable matching and a value
which is in reasonable agreement with the experimental value of 6.2.
It is dominated by the contributions from $Q_1$ and $Q_2$.
A large part
can be considered as the low-energy Penguins as we have discussed
in \cite{BPdIhalf}. For $G_{27}$ we do not have good stability.
The ENJL model improves very strongly on just using CHPT for the $X$-boson
couplings as used in
the work of the Dortmund group, but it still provides a too strong suppression.
The curve labelled ``I quad'' in Fig. \ref{figdIhalf} shows
the CHPT approximation for the $X$-boson couplings. The ENJL improvement
is obvious.

\begin{figure}
\includegraphics[height=0.49\textwidth,angle=270]{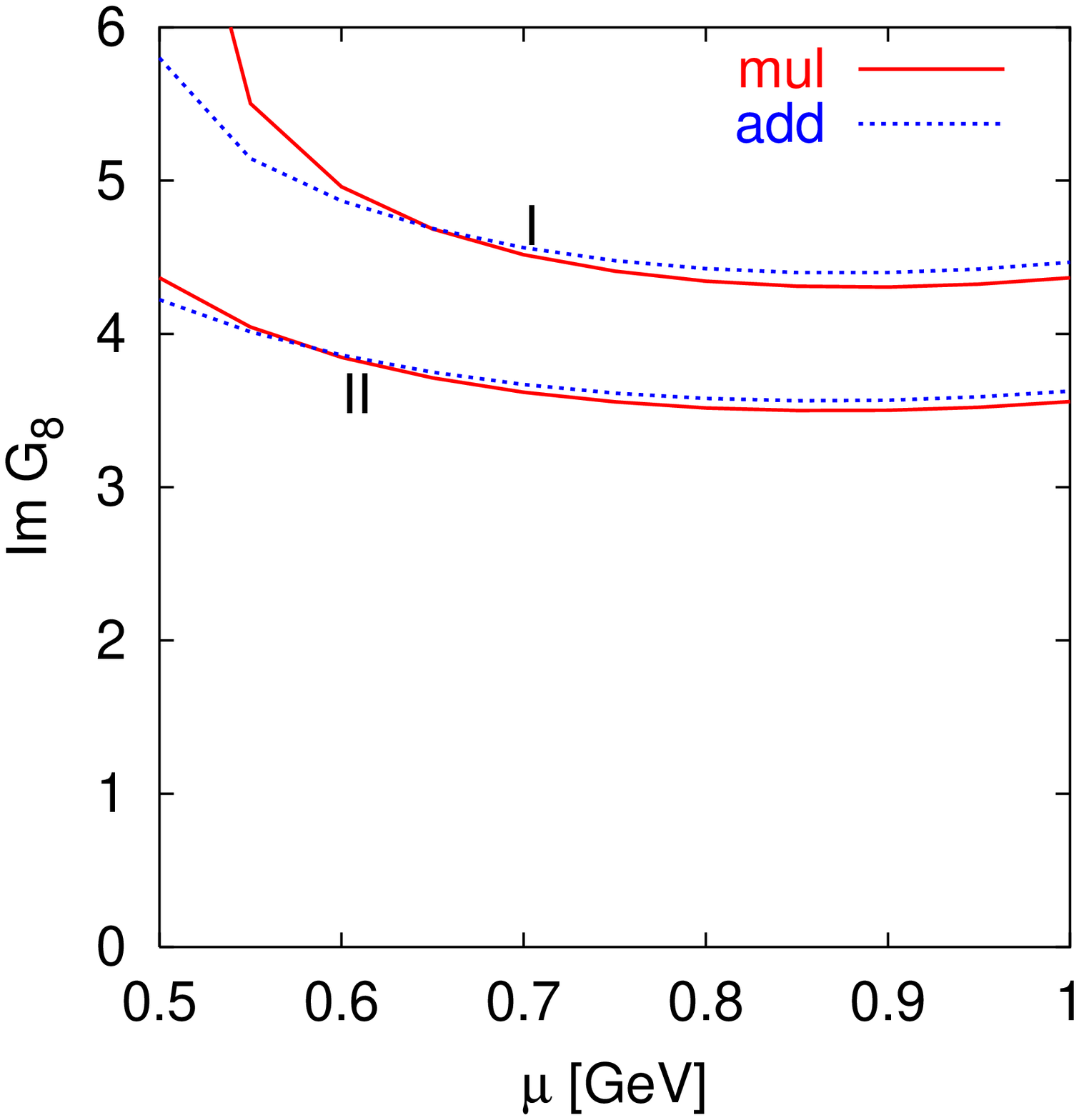}
\includegraphics[height=0.49\textwidth,angle=270]{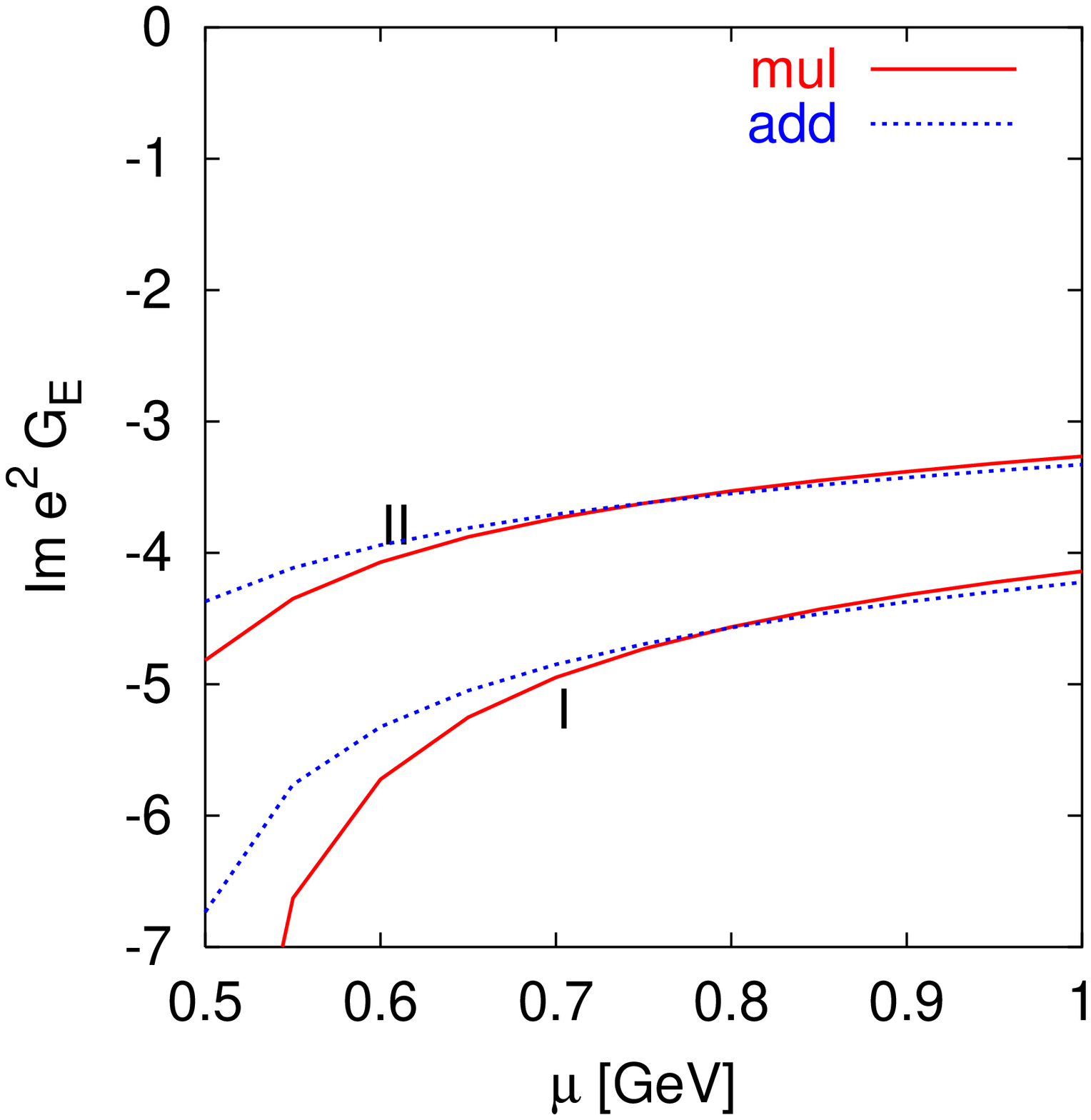}
\caption{\label{figImag} The part of $G_8$
and $e^2G_{E}$ multiplying $\tau$ as a function of $\mu$.
I and II refer to the two values of $\alpha_S$ used and the two curves
for each value of $\alpha_S$ are for
the two ways of solving the evolution equations as described in the text.}
\end{figure}

The parts multiplying $\tau$ of $G_8$ and $ e^2 G_E$ are shown
in Fig. \ref{figImag}. Notice that the strong contribution
proceeding mainly through $Q_6$ and dominating $G_8$ is from
the gluonic Penguin. The electroweak Penguins contribute via $e^2 G_E$
and the contribution is dominated by $Q_8$. The stability of both results
is quite acceptable. These numbers are the input used 
for $\varepsilon_K^\prime$ below.

\section{$B$-Parameters in the NDR Scheme}
\label{Bi}

The matrix-elements are usually quoted in terms of the bag-parameters.
These are defined as the ratio of the matrix-elements
divided by the vacuum-insertion-approximation (VIA) matrix-elements.
The latter have no scheme-dependence and so the bag-parameters are
also scheme-dependent. From our earlier results the bag-parameters
can be most easily determined in the $X$-boson scheme.
Afterwards they need to be converted to the more usual NDR-scheme.

The usual definitions as given in e.g. \cite{Buras,Peris} are
\ba
\label{B}
B_{1}^{(1/2)}(\mu) & \equiv & \frac{M_0[Q_1]}{\cal A}\;;
\nonumber\\
B_{2}^{(1/2)}(\mu) & \equiv  & -\frac{1}{5}\frac{M_0[Q_2]}{\cal A}\;;
\nonumber\\
B_{1}^{(3/2)} (\mu)& = 
& B_{2}^{(3/2)}(\mu) =  -\frac{\sqrt{2}}{8} \, \frac{M_2[Q_1]}{\cal A}\;;
\nonumber\\
B_{6}^{(1/2)} (\mu)& = & \frac{M_0[Q_6]}{\cal C}\;;
\nonumber\\
B_{7}^{(1/2)} (\mu)& = &  3 \frac{M_0[Q_7]}{\cal D}\;;
\nonumber\\ 
B_{7}^{(3/2)} (\mu) &=& 3 \sqrt{2} \frac{M_2[Q_7]}{\cal D}\;;
\nonumber \\
B_{8}^{(1/2)} ( \mu)& =  & \frac{M_0[Q_8]}{\cal D}\;.
\nonumber \\
B_{8}^{(3/2)} (\mu)&=&  \sqrt{2}\frac{M_2[Q_8]}{\cal D}\;;
\ea
with 
\ba
M_0[Q_j]&\equiv& \langle  (\pi \pi)_{I=0} | Q_j | K^0 \rangle \, , 
\nonumber \\
M_2[Q_j]&\equiv&
 \langle (\pi \pi)_{I=2} | Q_j | K^0 \rangle \, . 
\ea
and
\ba
{\cal A} &\equiv & -\frac{\sqrt 3}{9} F_0 \, (m_K^2 -m_\pi^2) 
\nonumber \\
{\cal C} &\equiv &  - 16  \sqrt 3 \, L_5 (M_\rho) \,
\frac{\langle 0 | \overline q q | 0 \rangle ^2(\mu)} {F_0^5}
(m_K^2-m_\pi^2)
\nonumber \\
{\cal D} &\equiv & - 2\sqrt 3  \, 
\frac{\langle 0 | \overline q q | 0 \rangle ^2(\mu)} {F_0^3}
\ea
We have restricted the definition of \cite{Buras} here to the lowest
order in CHPT for consistency. Full expressions away from the chiral limit are
in the first reference of \cite{epspeps}.

In the large $N_c$ limit one gets the model independent results \cite{BBG,BG87}
\ba
\label{BN_c}
B_{1}^{(1/2)} & = & 3\;;
\nonumber\\ 
B_{2}^{(1/2)} & = & \frac{6}{5} \;;
\nonumber\\
B_{1}^{(3/2)} & = & B_{2}^{(3/2)} \;=\;\frac{3}{4} \;;
\nonumber\\
B_{6}^{(1/2)} & = &  B_{8}^{(3/2)}  =  B_{8}^{(1/2)} = 1\;;
\ea
and
\ba
B_{7}^{(3/2)}  & = & B_{7}^{(1/2)}\; = 0\;.
\ea
Again these results are valid in the chiral limit. 
The operators $Q_7$ and $Q_8$ are discussed more below.

We normalized the $B_7$ parameter to the Fierzed part of the
vacuum insertion approximation.
This definition of the $B_7$ 
parameter is the same as
used by the lattice community \cite{latticeB7},
\cite{Peris} and in \cite{DG00}. Notice that in some cases terms beyond lowest
order are included in their definitions.

To lowest order in CHPT, i.e. ${\cal O } (e^0 p^2)$
or ${\cal O} (e^2 p^2)$ for the weak lagrangian one can rewrite the
definitions of Eq. (\ref{B})
in the notation used here and in \cite{BPdIhalf} for
the matrix-elements:
\ba
\label{Bchiral}
B_{1 \chi}^{(1/2)}(\mu) & = & - \frac{3}{5 C_1(\mu)}  \, 
(9 G_8[Q_1](\mu) + G_{27}[Q_1](\mu))\;;
\nonumber\\
B_{2\chi}^{(1/2)}(\mu) & = & \frac{3}{25 C_2(\mu)}  \, 
(9 G_8[Q_2](\mu) + G_{27}[Q_2](\mu))\;;
\nonumber\\
B_{1\chi}^{(3/2)}(\mu) & = & B_{2 \chi}^{(3/2)}(\mu) 
\;=\;\frac{3}{4 C_1(\mu)} \, G_{27}[Q_1](\mu)\;;
\nonumber\\
B_{6\chi}^{(1/2)}(\mu) & = &   - \frac{3 F_0^6}
{ 80 \langle 0 | \overline q q | 0 \rangle^2(\mu) L_5 (M_\rho)}\,  
\frac{G_8[Q_6](\mu)}{C_6(\mu) }\;;
\nonumber\\
B_{7\chi}^{(3/2)}(\mu) & = & B_{7\chi}^{(1/2)} (\mu)  =
-\frac{3 F_0^6}{5 \langle 0 | \overline q q | 0 \rangle^2(\mu)}
 \, \frac{e^2 G_E[Q_7](\mu)}{C_7(\mu)}\;;
\nonumber\\
B_{8\chi}^{(3/2)}(\mu) & = & B_{8\chi}^{(1/2)}(\mu) = 
-\frac{F_0^6}{5 \langle 0 | \overline q q | 0 \rangle^2(\mu)}
 \, \frac{e^2 G_E[Q_8](\mu)}{C_8(\mu)}\;.
\ea
The others can be derived similarly but we have only given the
numerically important ones and $B_7^{(3/2)}$ in order to compare
with other work \cite{Peris,latticeB7,DG00}.
 We have also restricted the VIA matrix-elements
to order $e^0 p^2$ and $e^2 p^0$ in order to be consistent with our full
calculation. 

Notice that even in the $X$-boson scheme the values of $B_{6}^{(1/2)}$, 
$B_7^{(I)}$, and $B_8^{(I)}$ depend on the value of $\alpha_S$ because of the
running of $B_0(\mu)$. The bag-parameters $B_6^{(I)}$ and $B_8^{(I)}$
the value at leading order in $1/N_c$
and the VIA value, run in the same way with $B_0(\mu)$ \cite{BG87}.

The  large $N_c$ result for the $Q_7$ operator 
is $O(e^2 p^2)$ in CHPT  and thus vanishes in the chiral limit.
The lowest $e^2 p^0$ order and numerically larger
contribution is NLO in $1/N_c$ and survives in the chiral limit.
This  operator together with $Q_8$ have been studied 
by several groups and techniques and deserves  special attention 
and will studied within the present approach in \cite{BGP}.

The numerical results for the bag-parameters in the $X$-boson
scheme can be easily derived
using the results from Section \ref{longdistance} and Ref. \cite{BPdIhalf},
in the case of lowest order CHPT couplings of the $X$-bosons
using Eqs. (\ref{geq8}) and (\ref{geq7}) and Eqs. (5.2)
and (5.17) of Ref. \cite{BPdIhalf}.
In the case of the ENJL model couplings for the $X$ bosons
one needs to use Eq. (\ref{geq8}), Table \ref{tabGEQ7}
and Tables 1, 2 and 3 of Ref. \cite{BPdIhalf}.
Notice that $B_8^{(I)X}= 1$ in the $X$-boson scheme even at NLO
in $1/N_c$ while
$B_6^{(I)X} = 2.2 \pm0.5$ \cite{PR99}.

\begin{figure}
\includegraphics[height=0.49\textwidth,angle=270]{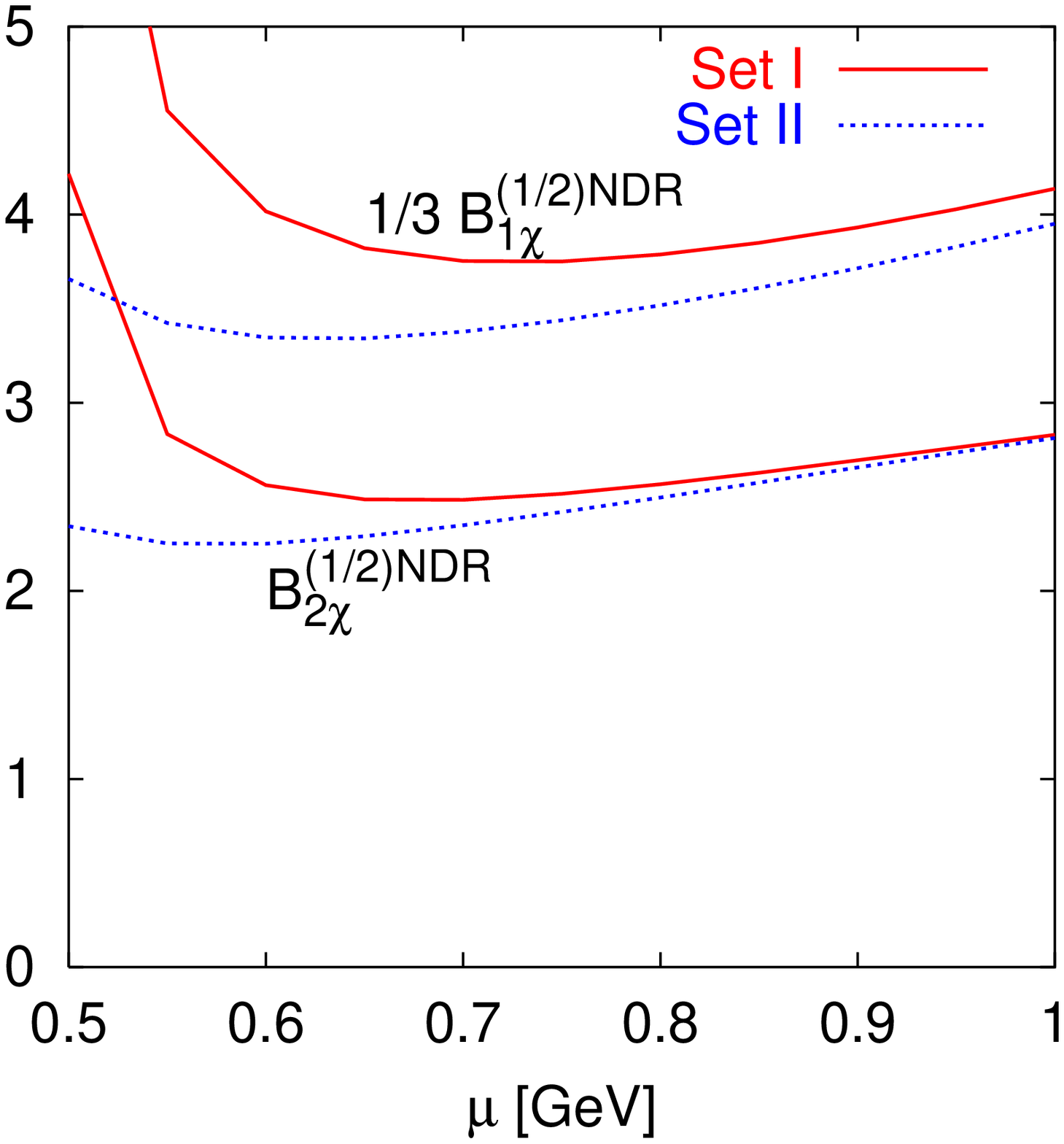}
\includegraphics[height=0.49\textwidth,angle=270]{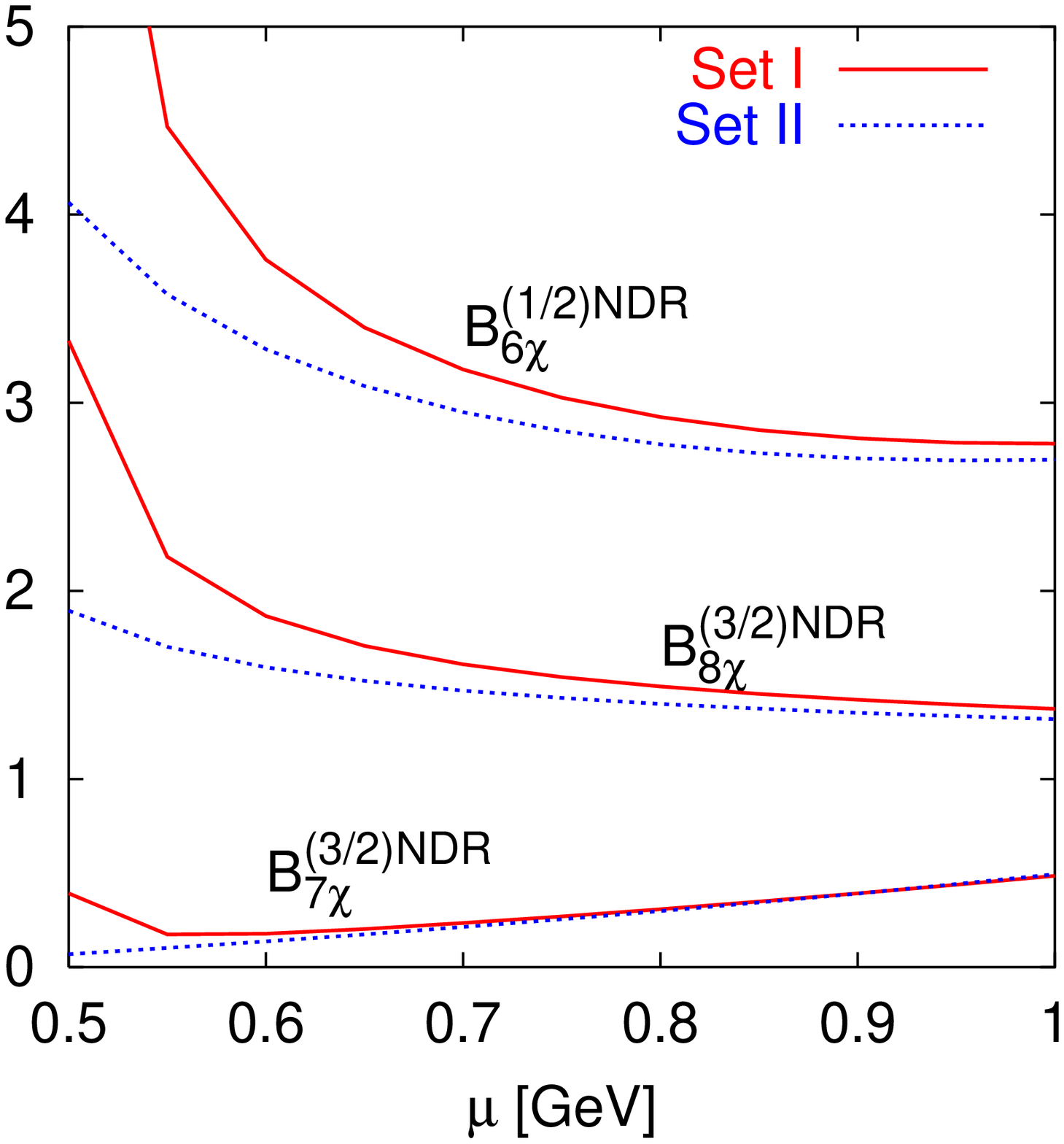}
\caption{\label{figBi} The bag parameters in the NDR
scheme for the two values of $\alpha_S$ we use. $B_{1\chi}^{(1/2)NDR}$
has been divided
by 3 to make it fit better in the plot. Notice the large values
for $B_{1\chi}^{(1/2)NDR}$ and $B_{2\chi}^{(1/2)NDR}$
responsible for the $\Delta I=1/2$ rule
and the large value for $B_{6\chi}^{(1/2)NDR}$
which is the reason for our large value
of $\varepsilon_K^\prime/\varepsilon_K$.}
\end{figure}

In order to get to the NDR-scheme we multiply with the combination
$r_{ij}-\tilde r_{ij}$ discussed in Section \ref{shortdistance} via
\be
\label{B_NDR}
B_{i}^{(I)NDR} = \frac{1}{\langle Q_i\rangle^I_{\mbox{\tiny VIA}}}
\sum_j \left[ \delta_{ij} -\frac{\alpha_S(\mu)}{\pi}
\left(r_{ij}-\tilde r_{ij}\right) -\frac{\alpha}{\pi}
\left(r_{ij}^e-\tilde r_{ij}^e\right)\right] B_{j}^{(I)X} 
\langle Q_j\rangle^I_{\mbox{\tiny VIA}}\;.
\ee
The $B_{j}^{(I)X}$-parameters are in the $X$-boson scheme.
We show the resulting numerical results for the inputs given earlier
in Fig. \ref{figBi}.
Notice the large values
for $B_{1\chi}^{(1/2)NDR}$ and $B_{2\chi}^{(1/2)NDR}$
responsible for the $\Delta I=1/2$ rule
and the large value for $B_{6\chi}^{(1/2)NDR}$
which is the reason for our large value
of $\varepsilon_K^\prime/\varepsilon_K$ while $B_{8\chi}^{(3/2)NDR}$
shows no such large enhancement. Notice though that we get
$B_{8\chi}^{(3/2)NDR}(\mu)
\approx 1.3 \pm 0.2$
 for $\mu$ between 0.6 GeV and 1 GeV which is around 30\% to
40\% larger than most other analysises.

In the NDR scheme and in the  chiral limit 
we get $B_{7 \chi}^{(3/2) NDR} (\mu) \approx 0.3 \pm 0.2$
for scales $\mu$ between 0.6 GeV and 1. GeV.  

\section{Results for $\varepsilon_K$ and $\varepsilon_K^\prime$ 
to Lowest Order}

The indirect CP-violation as described by $\varepsilon_K$, defined by
Eq. (\ref{defeps}), can be rewritten
to first order in CP violating parameters, using the $\Delta I=1/2$
rule and with $\re \, a_0 >> \varepsilon_K \, \im \, a_0$,
 to a very good approximation as
\ba
\label{defeps2}
\varepsilon_K &\simeq& \frac{e^{i \pi/4}}{\sqrt{2}}
\left(\frac{\im M_{12}}{\Delta m_K} + \frac{\im \, a_0}{\re \, a_0}\right)
\nonumber\\
M_{12}&=&\frac{G_F^2}{6\pi^2}f_K^2 \hat B_K m_K M_W^2\left[
(\lambda_c^*)^2 \hat \eta_1 S(x_c) + 
(\lambda_t^*)^2 \hat \eta_2 S(x_t) + 
2\lambda_c^* \lambda_t^* \hat \eta_3 S(x_c,x_t)\right] \, .
\ea
Putting\footnote{We use the experimental value of $G_{27}$ to
avoid the large $\mu$-dependence seen in our calculation of that quantity.
All other quantities were more stable as seen in Section \ref{resultsGi}.}
in the central numerical values of Section \ref{numinput},
$\sin(\delta)= 1$, $\re \, G_8 = 6.2$ and $G_{27}=0.48$
we obtain with $e^2 G_E$ and $\im \, G_8$ calculated in Section \ref{resultsGi}
\ba
\label{resulteps}
|\varepsilon_K| &=& (2.40-0.34) \cdot 10^{-3} = 2.06 \cdot 10^{-3}
\quad\mbox{Set I};\nonumber\\
|\varepsilon_K| &=& (2.49-0.28) \cdot 10^{-3} = 2.21 \cdot 10^{-3}
\quad\mbox{Set II}
\ea
at a value of $\mu = 0.75~GeV$. The $\mu$-dependence is very mild.
The first (second) number in brackets is the first (second) term in Eq.
(\ref{defeps2}).
The main difference is in the value of $\hat \eta_1$.
The effect from the second term which is more dependent on the quantities
calculated here, is small.

Given the uncertainty on $\hat B_K$ and the other input parameters this result
agrees well with the experimental value
\be
\label{epsexp}
\varepsilon_K^{\rm exp} = e^{i \Phi_{\varepsilon_K}} 
\, (2.280 \pm 0.013) \cdot 10^{-3} 
\ee
with $\Phi_{\varepsilon_K}= (43.49 \pm0.08)^0$.
For definiteness we will below use the value of Eq. (\ref{epsexp})
for our numerical estimates of $\varepsilon_K^\prime/\varepsilon_K$.

Using the same inputs we can now calculate $\varepsilon_K^\prime$
from its definition Eq. (\ref{defepsp}).
To first order in CP violating parameters, using the $\Delta I=1/2$
rule and with $\re \, a_0 >> \varepsilon_K \, \im \, a_0$, we can
use to a very good approximation
\be
\label{defepsp2}
\varepsilon^\prime_K \simeq
\frac{1}{\sqrt{2}}\frac{\re \, a_2}{\re \, a_0}
\left(-\frac{\im \, a_0}{\re \, a_0}+\frac{\im \, a_2}{\re \, a_2}\right)
\, e^{i \Phi_{\varepsilon_K'}}\, .
\ee
With $\Phi_{\varepsilon_K'}= \pi/2 + \delta_2 -\delta_0
= (48 \pm 4 )^0$ the strong phase.

\ba
\label{resultepsp}
\left|\frac{\varepsilon_K^\prime }{\varepsilon_K}\right| &=&
 (9.2-2.5) \cdot 10^{-3} = 6.7  \cdot 10^{-3}
\quad\mbox{Set I}\nonumber\\
\left|\frac{\varepsilon_K^\prime }{\varepsilon_K}\right| &=&
 (7.4-1.9) \cdot 10^{-3} = 5.5  \cdot 10^{-3}
\quad\mbox{Set II}
\ea
at a value of $\mu=0.75~GeV$.
The first (second) number in brackets is the first (second) term in Eq.
(\ref{defepsp2}).
The first number is dominated by $Q_6$ and is usually called the strong 
Penguin contribution.
The second one is from the contribution
of $Q_7$ to $Q_{10}$ and is dominated by $Q_8$ and is usually called
the electroweak Penguin contribution.

\begin{figure}
\includegraphics[height=0.9\textwidth,angle=270]{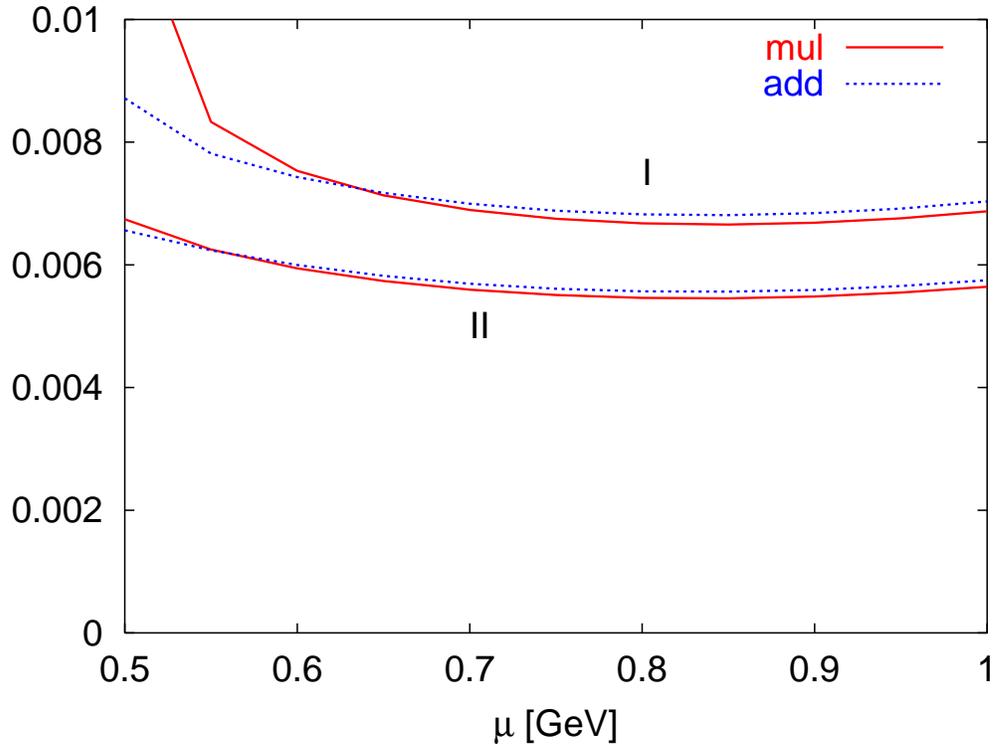}
\caption{\label{figepspeps}
Our result for $\varepsilon_K^\prime/\varepsilon_K$ in the chiral limit
using the experimental values of $\varepsilon_K$ and the inputs as described in
the text.
I and II refer to the two values of $\alpha_S$ used and the two curves
to two ways of solving the evolution equations as described in the text.}
\end{figure}

We have also shown the $\mu$-dependence of the final result
for the two values of $\alpha_S$ and the two ways of NLO resumming
used in Fig. \ref{figepspeps}.
The predictions are quite stable, partly due to the fact that
we fully calculate all contributions in the same way.

Our final prediction to lowest order in CHPT is 
\ba
\label{resultepsploworder}
\left| \frac{\varepsilon_K^\prime}{\varepsilon_K} \right|_{O(p^2)}
&=& (6\pm3) \cdot 10^{-3} \, 
\ea
We used the experimental value of $\varepsilon_K$ in the above numbers
but using the values we calculated would not change any results
dramatically.

The error we quoted there rest on several things,
the matrix-elements of $Q_6$ and $Q_8$ depend directly
on $B_0^2(1~{\rm GeV})$, this does not affect the relative size
of $Q_6$ and $Q_8$ but
does affect the overall value in addition to
the typical 30\% or so we expect our method to have; we did not include
any uncertainties in CKM matrix elements and as discussed below, there are
several more effects which we did not include yet.

\section{Discussion and Corrections}

Our number is significantly higher than any others discussed in
the recent literature\cite{Burasetc}. 
The strong Penguin in our case is enhanced in two ways compared
to the usual treatment. First the scheme where the transitions to models
can be made must be taken into account. This provides a first enhancement.
Then in the $X$-boson scheme and in the chiral limit
the $B_6$-parameter takes a value of
$B_{6 X}^{\chi}\approx 
2.2\pm0.5$ \cite{PR99} and $B_{8 X}^{\chi} = 1$
at next-to-leading order in $1/N_c$. Putting it together with the first
effect leads to $B_{6 \chi}{\mbox{\tiny NDR}}(\mu) \approx2.9 \pm 0.7$ 
in the chiral limit for scales between 0.6 GeV  and 1 GeV as discussed above.
We do not see any similar enhancement for 
$B_{8\chi}^{\mbox{\tiny NDR}}(\mu) \approx 1.3 \pm 0.2$,   
so the usual
strong cancellation between the strong and electroweak Penguins is much weaker
in our calculation. A last reason is that we treat both the real
and imaginary part at the same order, not as is usually done by treating
the real part to all orders using experiment and the imaginary part
to lowest order.

To the value above one has to add higher order CHPT corrections
which are proportional to pion and Kaon masses. These are of three
types, namely:
(1) final state interactions (FSI) from $\pi-\pi$ interactions
(2) CHPT corrections which are purely
real, they are due to $\Delta S=1$ higher order terms in the Lagrangian
 and real loop-diagrams;
(3) isospin breaking corrections due to quark masses, usually estimated
by $\pi_0-\eta$ mixing, and electromagnetic corrections.

FSI have  recently been  considered in Ref. \cite{Pallante} and later also
in \cite{BurasFSI}. The main point of \cite{Pallante} is that
both the real and the imaginary part of the amplitudes should be treated
in the same way leading to a significant enhancement over the usual
estimates. We already treat the imaginary and the real parts
in Eq. (\ref{defepsp2}) in the same way so the main effect of FSI
is in changing the ratio $\re \, a_0/\re \, a_2$ from about 16.2 to 22.2.
Including only FSI to {\em all orders} then changes Eq. (\ref{resulteps})
to
\ba
\label{resultepspFSI}
\left|\frac{\varepsilon_K^\prime }{\varepsilon_K}\right| &=&
 (6.8-1.8) \cdot 10^{-3} =  5.0 \cdot 10^{-3}
\quad\mbox{Set I}\nonumber\\
\left|\frac{\varepsilon_K^\prime }{\varepsilon_K}\right| &=&
 (5.4-1.4) \cdot 10^{-3} =  4.0  \cdot 10^{-3}
\quad\mbox{Set II}
\ea

Isospin breaking effects due to
the $\pi^0-\eta$ mixing to $O(p^4)$ have been calculated 
in Ref. \cite{EMNP} and references therein.
This effect adds  a contribution  to  $\im \, a_2$  which is
parameterized usually as
\ba
\frac{\left[\im \, a_{2}\right]_{IB}}{\re \, a_2} &=& \Omega_{IB} \, 
\frac{\im \, a_0}{\re \, a_0} \, . 
\ea
Using  the value $\Omega_{IB} = 0.16\pm 0.03$ \cite{EMNP}
one gets
\ba
\label{resultepsfull}
\left|\frac{\varepsilon_K^\prime }{\varepsilon_K}\right| &=&
 (6.8-2.9) \cdot 10^{-3} =  3.9 \cdot 10^{-3}
\quad\mbox{Set I}\nonumber\\
\left|\frac{\varepsilon_K^\prime }{\varepsilon_K}\right| &=&
 (5.4-2.3) \cdot 10^{-3} = 3.1  \cdot 10^{-3}
\quad\mbox{Set II}
\ea

Corrections of type (2) have been fitted to $O(p^4)$ in 
 \cite{Kambor} and calculated to the same order 
in \cite{Bertolini} but disentangling
this from their FSI effects is not obvious. Notice that
in \cite{Dortmund} and \cite{Bertolini} the enhancement found for
$B_6$ is from this source and not from the one discussed here.

Isospin breaking corrections beyond mixing have been discussed recently
as well, see  \cite{GV99}, but their full estimate is not done at
the present stage.
Our approach allows to include them in a more systematic fashion
and this will be done in the future.

Given the uncertainties on the CKM matrix-elements which we did
not include and the uncertainties on the hadronic matrix elements
which we still face, a reasonable guess at the final result
is
\be
\left|\frac{\varepsilon_K^\prime}{\varepsilon_K}\right| =
 (3.4\pm1.8)\cdot 10^{-3}
\,.
\ee
This  result is somewhat above the
present world average \cite{experiment}
\be
\left|\frac{\varepsilon_K^\prime}{\varepsilon_K} \right|^{\rm exp}
= (2.13 \pm 0.46) \cdot 10^{-3}\,
\ee
but quite compatible with it.

In conclusion
we have calculated the chiral limit prediction for
$\varepsilon_K^\prime/\varepsilon_K$ in a consistent fashion using the $1/N_c$
expansion and the $X$-boson method. The result is substantially
larger than most other estimates. After including the two largest
expected corrections we obtain a result in acceptable agreement with
the experiment.

\section*{Acknowledgements}
This work has been partially supported  by the Swedish Science Foundation.
The work of J.P. was supported in part by CICYT (Spain) and by
Junta de Andaluc\'{\i}a under Grants Nos. AEN-96/1672 and FQM-101
respectively. 
J.P. thanks Matthias Jamin for correspondence on the singularity
cancellations in the Wilson coefficients and discussions with Toni Pich.
He also likes to thank the Department of Theoretical Physics at Lund 
University (Sweden) where part of his work was done for hospitality.
We thank Elisabetta Pallante for participation  in the early stages
of this work.

\appendix
\section{Scheme Dependence}

The matching conditions of Eq. (\ref{match2})
between  the effective Lagrangian action (\ref{effactionOPE}) and the effective
$X$-boson (\ref{effactionX}) in 
Section \ref{shortdistance}  include  two ten by ten matrices. 
The explicit calculation gives for the gluonic corrections
\ba
\lefteqn{ 
r -\tilde r =\frac{1}{4}\times }&&\nonumber\\&&
\hskip-16pt\left(
\begin{array}{cccccccccc}
\frac{11}{2 N_c} & \frac{-11}{2} &  0 &  0 &  0 &  0 &  0 &  0 &  0 &  0  
\\[1mm]
\frac{-11}{2} & \frac{11}{2 N_c} &  0 &  0 &  0 &  0 &  0 &  0 &  0 &  0  
\\[1mm]
0 & \frac{-7}{18 N_c} &  \frac{85}{18 N_c} & -\frac{99 N_c + n}{18 N_c}
& 0 & \frac{4 n}{9 N_c} & 0 & \frac{4 \tilde n}{9 N_c} & 
\frac{7}{18 N_c} & - \frac{\tilde n}{18 N_c} 
\\[1mm]
0 & \frac{7}{18} & \frac{85}{18} & \frac{99 + n N_c }{18 N_c} & 0
& -\frac{4 n}{9} & 0 & -\frac{4 \tilde n}{9} & -\frac{7}{18} &
\frac{\tilde n}{18}
\\[1mm]
0 & -\frac{7}{18 N_c} & - \frac{7}{9 N_c} & -\frac{n}{18 N_c} &
\frac{1}{2 N_c} & \frac{-27 N_c + 4 n}{ 9 N_c} & 0 & 
\frac{4 \tilde n}{9 N_c} & \frac{7}{18 N_c} &
-\frac{\tilde n}{18 N_c} 
\\[1mm]
0 & \frac{7}{18}& \frac{7}{9} & \frac{n}{18} & -\frac{1}{2} &
\frac{27-4 n N_c}{9 N_c} + 5 C_F & 0 & -\frac{4 \tilde n}{9} &
-\frac{7}{18} & \frac{\tilde n}{18} 
\\[1mm]
0 & 0 & 0 & 0 & 0 & 0 & \frac{1}{2 N_c} & -3 & 0 & 0
\\[1mm]
0 & 0 & 0 & 0 & 0 & 0 & -\frac{1}{2} & \frac{3}{N_c} + 5 C_F &
0 & 0
\\[1mm]
0 & 0 & 0 & 0 & 0 & 0 & 0 & 0 & \frac{11}{2 N_c} & -\frac{11}{2}
\\[1mm]
0 & 0 & 0 & 0 & 0 & 0 & 0 & 0 & -\frac{11}{2 } & \frac{11}{2 N_c}
\end{array}
\right) \nonumber \\
\ea
with 
\be
C_F = \frac{N_c^2-1}{2 N_c} \, . 
\ee

The photonic corrections are
\ba
\lefteqn{
r^e -\tilde r^e =
\frac{1}{4} \times}&&\nonumber\\&&
\hskip-20pt\left(
\begin{array}{cccccccccc}
\frac{22}{9} & 0 & 0 & 0 & 0 & 0 & 0 & 0 & 0 & 0
\\[1mm]
0 & \frac{22}{9} & 0 & 0 & 0 & 0 & 0 & 0 & 0 & 0
\\[1mm]
0 & 0 & 0 & 0 & 0 & 0 & 0 & 0 & \frac{11}{9} & 0
\\[1mm]
0 & 0 & 0 & 0 & 0 & 0 & 0 & 0 & 0 & \frac{11}{9}
\\[1mm]
0 & 0 & 0 & 0 & 0 & 0 & \frac{1}{9} &0 & 0 &0
\\[1mm]
0 & 0 & 0 & 0 & 0 & 0 & 0 & \frac{1}{9} & 0 & 0
\\[1mm]
\frac{4 N_c}{81} & \frac{28}{81} & \frac{-28- 20 \tilde n N_c}{81} &
\frac{-4N_c +4 \tilde n}{81} & \frac{18-20\tilde n  N_c}{81} &
 \frac{-32 \tilde n}{81} & \frac{9-20 \overline{n} N_c}{81} &
 \frac{-32 \overline{n}}{81} & \frac{14-20 \overline{n} N_c}{81} &
 \frac{2 N_c + 4 \overline{n}}{81}
\\[1mm]
0 & 0 & 0 & 0 & 0 & \frac{2}{9}  & 0 & \frac{1}{9} &0 &0
\\[1mm]
\frac{4 N_c}{81} & \frac{28}{81} & \frac{22}{9} & 
 \frac{-4N_c+4\tilde n}{81} & \frac{-20 \tilde n N_c}{81} &
\frac{-32\tilde n}{81} & \frac{-20 \tilde n N_c}{81} &
\frac{-32 \overline{n}}{81} & \frac{113-20 \overline{n}N_c}{81} &
\frac{2 N_c + 4 \overline{n}}{81} 
\\[1mm]
0 & 0 & \frac{-28-20 N_c \overline{n}}{81} & \frac{22}{9} &
0 & 0 & 0 & 0 & 0 & \frac{11}{9} 
 \end{array} 
\right) \nonumber \\
\ea
with
\be
n = n_u + n_d;\quad
\tilde n = n_u -\frac{n_d}{2}\quad\mbox{and}\quad
\overline{n} = n_u + \frac{n_d}{4}
\ee
and $n_u$ the number of up-like light-quarks
and $n_d$ the number of down-like light-quarks.

\end{document}